\documentclass[fleqn,usenatbib]{mnras}
\usepackage{longtable}
\usepackage{lipsum}
\usepackage{newtxtext,newtxmath}
\usepackage[T1]{fontenc}

\DeclareRobustCommand{\VAN}[3]{#2}
\let\VANthebibliography\thebibliography
\def\thebibliography{\DeclareRobustCommand{\VAN}[3]{##3}\VANthebibliography}

\usepackage{graphicx}	% Including figure files
\usepackage{amsmath}	% Advanced maths commands

\title[GW170817 Afterglow \& Jet Dynamics]{The Late-time Afterglow of GW170817 and Implications for Jet Dynamics}

\author[Katira et al.]{
{Aman Katira },$^{1,2}$\thanks{E-mail: Amankattira11@gmail.com}
Kunal~P.~Mooley,$^{1,3}$
Kenta Hotokezaka$^{4}$
\\
$^{1}$Indian Institute Of Technology Kanpur, Kanpur, Uttar Pradesh 208016, India\\
$^{2}$Indian Institute Of Technology Mandi, Kamand, Himachal Pradesh 175075, India\\
$^{3}$Caltech, 1200 E. California Blvd. MC 249-17, Pasadena, CA 91125, USA\\
$^{4}$Research Center for the Early Universe, Graduate School of Science, The University of Tokyo, Bunkyo, Tokyo
}

\date{Accepted XXX. Received YYY; in original form ZZZ}

\pubyear{\the\year{}}

\begin{document}
\label{firstpage}
\pagerange{\pageref{firstpage}--\pageref{lastpage}}
\maketitle

% Abstract of the paper
\begin{abstract}
GW170817 is the first binary neutron star merger detected with gravitational and electromagnetic waves, and its afterglow is still detectable 7 years post-merger.  
Some previous studies of the X-ray afterglow have claimed the onset of a new afterglow component or raised concerns about the data processing techniques. 
Motivated thus, we present here a reanalysis of X-ray afterglow data for GW170817 and find potential sources of discrepancies between the data reduction techniques employed by various research groups.
We also analyze the updated panchromatic afterglow data to find that there is no significant evidence for any new afterglow component (e.g. due to the ejecta that gave rise to the kilonova) and that the jet must be still in a mildly relativistic phase.
 The decline in the afterglow light curve is significantly shallower compared to that expected from the standard synchrotron afterglow jet models with sideways spreading, indicating either an additional energy injection at late times or the velocity dependence on the microphysics parameters. In this context, we discuss the implications of the late time afterglow data on jet dynamics. 
\end{abstract}

% Select between one and six entries from the list of approved keywords.
% Don't make up new ones.
\begin{keywords}
catalogs --- X-ray: neutron stars, gravitational waves
\end{keywords}

%%%%%%%%%%%%%%%%%%%%%%%%%%%%%%%%%%%%%%%%%%%%%%%%%%

%%%%%%%%%%%%%%%%% BODY OF PAPER %%%%%%%%%%%%%%%%%%

\section{Introduction}

The gravitational-wave (GW) multi-messenger discovery of the binary neutron star merger GW170817 \citep{GW170817-MMA}, and the extensive electromagnetic (EM) data gathered for this event, gave unique insights into open questions in fundamental physics, astrophysics, chemistry and cosmology.
The GW data primarily constrained the masses of the component neutron stars and remnant and tidal deformability \citep{GW170817}. 
% \muskan{please see if this paper is valid here- A constraint on the dissipative tidal deformability of neutron stars. It's Ripley2024}

The gamma-ray pulse that arrived 1.7 s after the merger \citep{Savchenko2017,Fermi2017} was found to originate in mildly relativistic outflow ($\Gamma\sim 5$; referred to as "cocoon" or "structured jet wing" in literature) that left its signatures in the afterglow signal within the first 150 days post-merger \citep{hallinan2017,mooley2018-wideoutflow,Matsumoto2019,nakar2020,Troja2017,Lazzati2018,lyman2018,Lamb2019,DAvanzo2018,Xie2018,Margutti2018,Resmi2018,Nynka2018,corsi2018}. The combination of GW and gamma-ray signals constrained the speed of gravity, gave new bounds on the violation of Lorentz invariance and facilitated a new test of the equivalence principle \citep{GW170817-GW-Gamma}.

The kilonova/macronova, a thermal transient born from the heating due to r-process nucleosynthesis \citep{andreoni2017follow,arcavi2017,buckley2018,chornock2017,coulter2017,cowperthwaite2017,Diaz2017,Drout2017,Evans2017,Hu2017,Kasliwal2017, Kilpatrick2017,Lipunov2017, McCully2017, Nicholl2017, Pian2017, Pozanenko2018, Shappee2017, Smartt2017, soares-santos2017, Tanvir2017, Tominaga2018, Utsumi2017, Valenti2017, Villar2017, Villar2018, margutti-chornock2021, kasliwal2019}, dominated the UV-optical-near infrared part of the EM spectrum between $\sim0.5-30$ days post-merger. It offered a rare opportunity to study the formation of very heavy elements ($Z\gtrsim50$) and revealed a $v\sim0.1-0.3$c outflow.

Due to the proximity of the event \citep[localized to NGC 4993 at 40 Mpc; e.g.][]{coulter2017,Kasliwal2017}, high-angular resolution measurements, carried out with the Hubble Space Telescope \citep{Mooley2022} and radio very long baseline interferometry \citep{mooley2018-vlbi,ghirlanda2019}, yielded precise measurements of superluminal motion and constraint on the compact size of the radio source. This, together with the narrow peak and rapid decline ($F_\nu \propto t^{-1.8}$) of the panchromatic afterglow light curve \citep{ddobie2018,troja2022,mooley2018-strongjet} gave credence to a narrow and highly relativistic jet core. The proper motion of the jet set an upper limit on the initial lorentz factor of the jet core: $\Gamma > 40$ \citep[90\% confidence;][]{Mooley2022}.
The consensus picture on the afterglow of GW170817 is therefore that of a short GRB where the jet was seen off-axis, at an angle of about $20^o$ \citep{fong2017,Mooley2022,alexander2018,margutti-chornock2021}, and for the first time we were able to probe the angular structure of the jet\footnote[1]{Described as a highly relativistic jet core surrounded by a mildly relativistic cocoon, or in other words, a structured jet having highly relativistic core and mildly relativistic wings.}.
The GW and EM data, together with the precise constraints on the viewing angle obtained from the afterglow light curve and proper motion measurements, gave a precise standard-siren measurement \citep{Schutz1986} of the  Hubble-Lemaitre constant in the local universe \citep{hotokezaka2019,Mooley2022}. 

While the afterglow of GW170817 has now, 7.5 years post-merger, faded below the sensitivity limits of most current telescopes \citep{Kilpatrick2022, Balasubramanian2021, Balasubramanian2022}, it is still detectable with deep Chandra observations \citep[][PI Troja]{ryan2024}.
The late-time afterglow emission has been of much interest to the astronomical community given the late-time rebrightening expected \citep{nakar2011,hotokezaka2015} from the non-relativistic outflow that dominated the kilonova emission.
Recent works have claimed X-ray excess/flattening and discrepancies in the data processing methods of various research groups \citep{troja2022,ryan2024,Makhathini2021,hajela2022}.
Here we reprocess the X-ray dataset and use updated panchromatic afterglow data to 1) test whether the jet is still in the relativistic phase, 2) find whether there is any excess afterglow emission at late time (7 years post-merger; that may be indicative of a new afterglow component), and 3) investigate the implications of GW170817's late time afterglow data on jet dynamics. 
In this paper we present the reanalysis of X-ray data in Sec.~\ref{sec:data}, compilation and model fitting of the updated pachromatic afterglow data in Sec.~\ref{sec:panchromatic} and discussion in Sec.~\ref{sec:discussion}.

\begin{figure*}
\includegraphics[height=0.35\textheight]{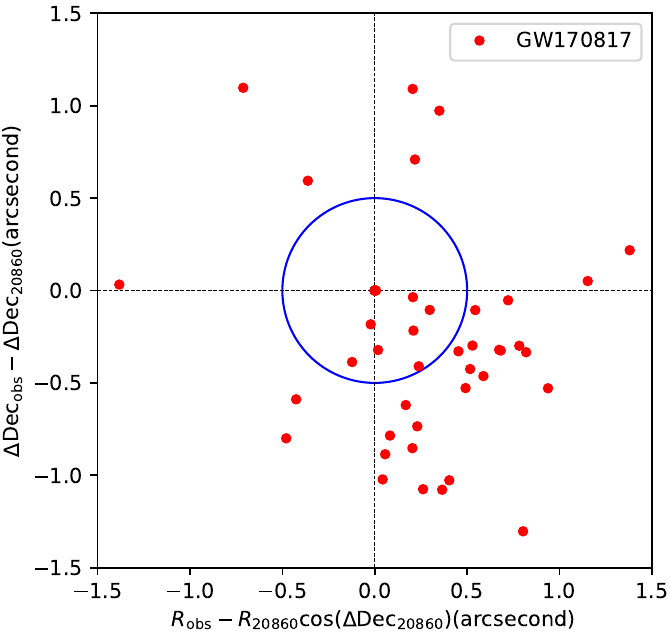} % Adjust height
\includegraphics[height=0.35\textheight]{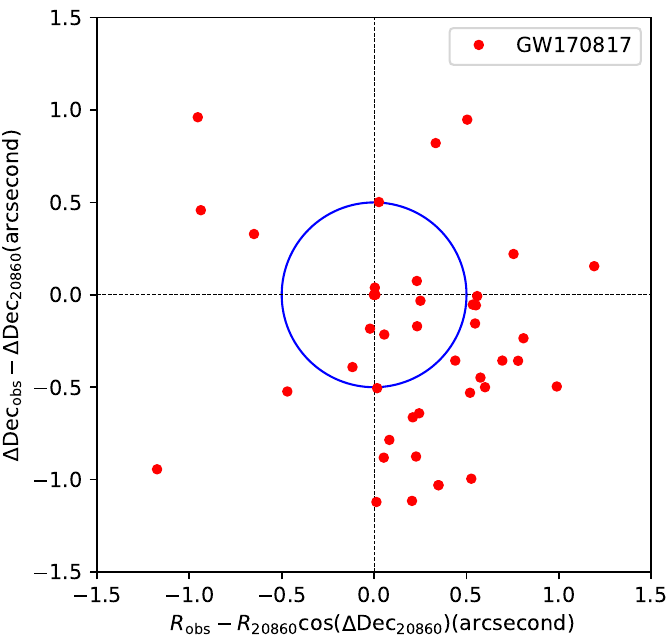} % Adjust   
\includegraphics[height=0.35\textheight]{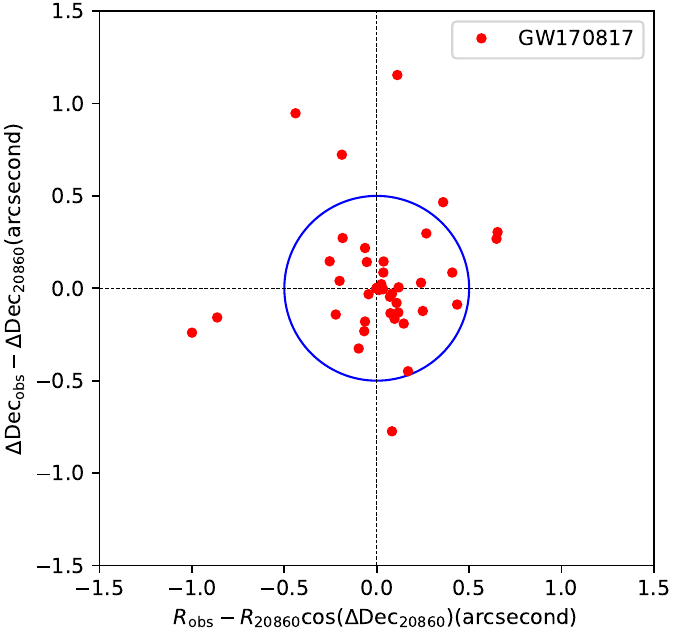} % Adjust height
\includegraphics[height=0.35\textheight]{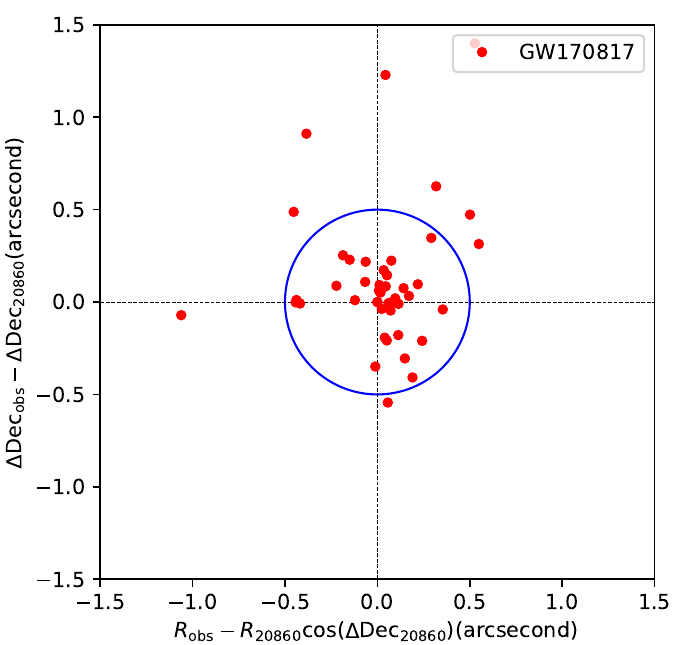} % 
\caption{The distribution of positional offsets in the right ascension (RA) vs. declination (Dec) plane for GW170817. The offsets are shown for all observations with respect to the source position in ObsID 20860, after applying two different methodologies.
1. Updating the aspect solution file using \textit{"Repro\_aspect"} (Upper Left), and then using that aspect solution file to reproject the observation in a single epoch to a common tangent point using \textit{"Reproject\_obs"} (Upper Right).
2. Updating the header of the event file using \textit{"wcs\_match"} and \textit{"wcs\_update"} (Lower Left), and then using the updated event file to reproject the observation in a single epoch to a common tangent point using \textit{"Reproject\_obs"} (Lower Right). Due to the much tighter alignment on the position of GW170817 with the \textit{wcs\_match} \& \textit{wcs\_update} method, we decided to use this method to calculate the X-ray fluxes.}
% \muskan{which method among these did we use and why? -- please clarify this in the figure caption as well as in the main text}} \myadav{I added in text.}
\label{fig:positional_offset}
\end{figure*}

\begin{figure}
    \centering
    \includegraphics[width=0.5\textwidth]{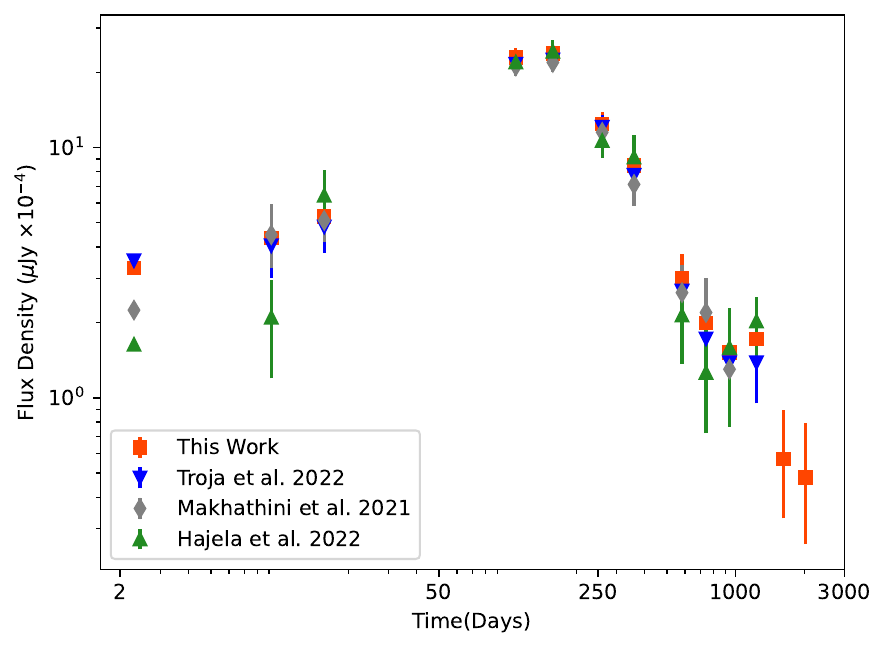}
    \caption{The X-ray flux density (at $\nu=2.4\times10^{17} \equiv$ 1 keV) light curve for the afterglow of GW170817. The values from this work together with those published in literature \citep{Makhathini2021,troja2022,hajela2022} are shown. The flux density values at 2.3 days correspond to 3$\sigma$ upper limits. See Sec.~\ref{sec:data} for more details.} 
    \label{fig:lc-2methods}
\end{figure}

\begin{figure}
    \includegraphics[width=3.5in]{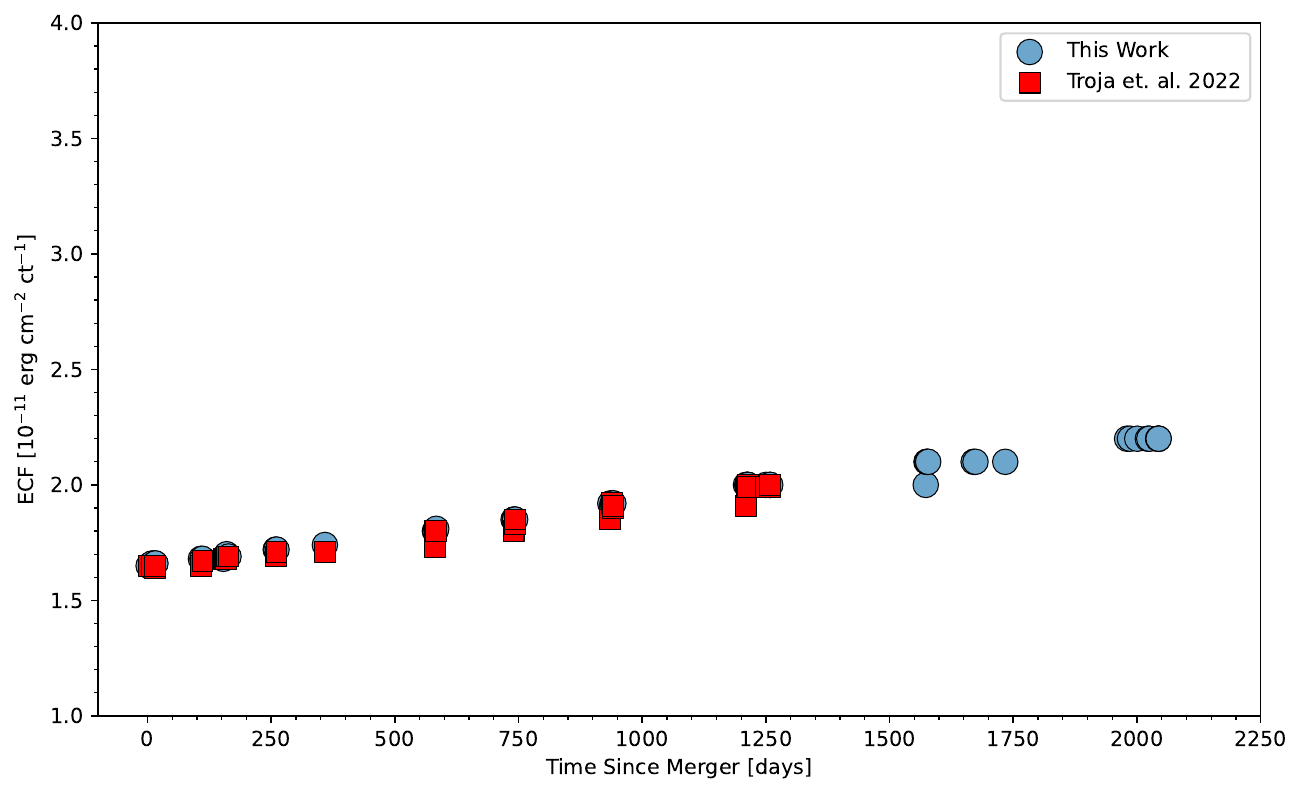}
    \caption{The Energy Conversion Factor (ECF) derived for all the GW170817 data taken with the Chandra X-ray Observatory is a smoothly rising function as expected. The ECFs from this work are consistent with those from \citep{troja2022}; any small differences are likely due to the updated CALDB files in the latest version of CIAO that we have used in this work.}
    \label{fig:ECF}
\end{figure}

\begin{figure}
    \centering
    \includegraphics[width=0.5\textwidth]{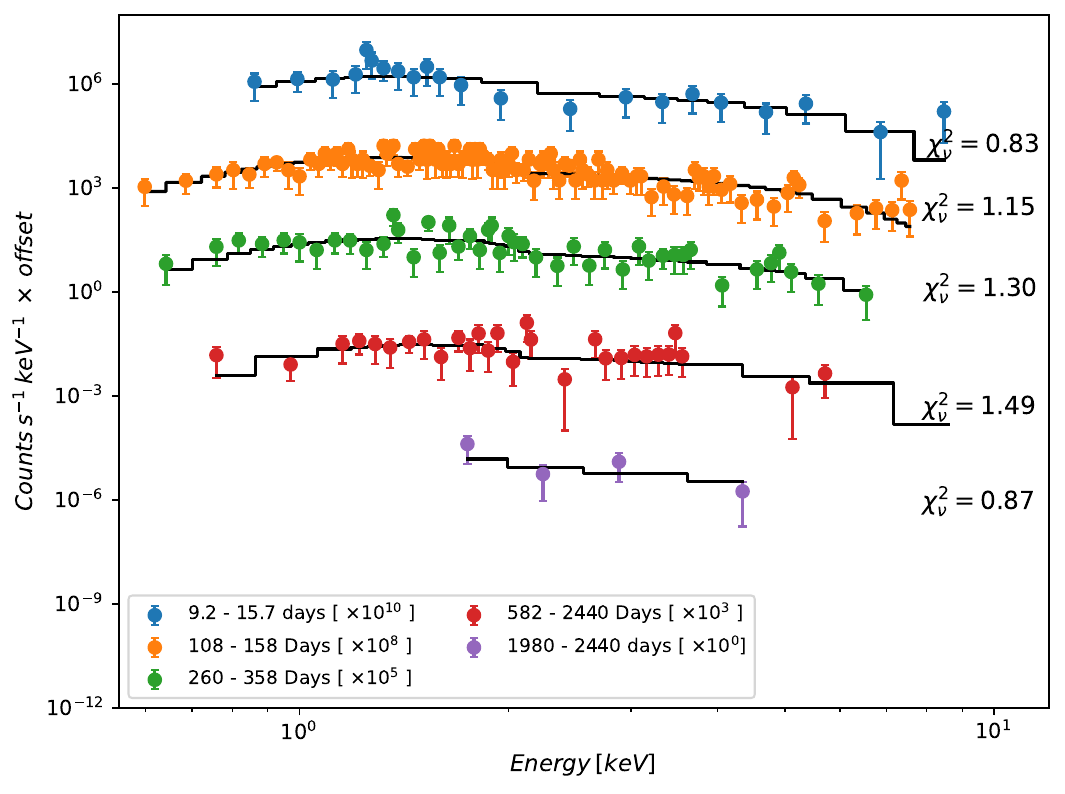}
    \caption{X-ray spectra of GW170817 at various epochs (color coded). A power law (black solid line) with spectral index $\beta = -0.58$ (Photon index = 1.58) gives good fits to the spectra at all epochs ( $\chi_\nu^2 \simeq 1$). The lack of any evolution in the spectral index, at especially the later epochs, indicates that the cooling break has likely not passed through the X-ray band $\sim$4--6 years post-merger and that a single component (jet afterglow) satisfactorily explains the late-time afterglow without the need for invoking alternate emission mechanisms (e.g. afterglow from the kilonova ejecta). Epoch 14 refers to observations carried out across Apr-May 2024.}
    \label{fig:xray-spec}
\end{figure}

\section{Reanalysis of the Chandra X-ray Data}\label{sec:data}
The Chandra X-ray Telescope observed GW170817 between August 19, 2017 (2.3 days post-merger) and March 23, 2023 (2043 days post-merger). We have reprocessed these data, consisting of 47 observations grouped into 13 epochs (closely-spaced observations), using CALDB 4.11.0 calibration files and CIAO v. 4.16. A log of individual observations and epochs is given in Tables~\ref{tab:flux_density} and~\ref{tab:Individual_obs}.

Using the \textit{chandra\_repro} tool, we processed the data as per standard procedure.  
Every exposure was then matched to a common frame, which was essential for identifying X-ray emission from GW170817 across several observations. Using wavelet sizes of 1, 2, and 4 pixels, we used \textit{wavdetect} to locate X-ray sources in the 0.5-7 keV range. To align astrometry of each obsID to a reference frame (obsID 20860), one can follow two procedures. One, by using \textit{repro\_aspect} which updates the aspect files and the new aspect file can then be used for reprojection. Second, is by using \textit{wcs\_match} which filtered sources within 1'' and a residual limit of 1'' and \textit{wcs\_update} to reprocess every observation and updated the astrometry which modifies the event files' headers. \textit{deflare} was utilised to eliminate background flares, employing a sigma clipping threshold of 3. Using \textit{reproject\_obs} which basically reprojects these observations to a common tangent point in order to enhance the alignment between diverse observations. We combined observations made at comparable epochs into a single image. We show the distribution of positional offsets, in the right ascension vs declination plane, across individual exposures and approaches (outlined above) in Figure~\ref{fig:positional_offset}. This figure demonstrates that the aligning of points degree to which different approaches align the observations. 
This is likely a source of discrepancy between the Chandra data processing carried out by previously by different research groups \citep{troja2022,Makhathini2021,hajela2022}.
We find a much tighter alignment on the position of GW170817 with the \textit{wcs\_match} \& \textit{wcs\_update} approach compared with the \textit{reproject\_obs} approach, and therefore we use the former approach to calculate the flux and flux density values (described below). 
Nevertheless, when we calculated the flux density values we generally find less than 5\% difference between the two methods except for 2 epochs where the difference is $\sim$20\%.
% \myadav{Please make it better :)}

To find the count rates and fluxes we conducted aperture photometry (procedure described below) in the energy range of 0.5-7.0 keV. Aperture radius of $1.5''$ (encompassing 92\% of the point spread function at 2.3 keV, as calculated using the \textit{psfsize\_srcs} task) was used for observations with more than 15 source counts, and $1''$ (encompassing 90\% of the PSF) for observations with fewer counts in order to optimize the signal-to-noise ratio and avoid contamination from neighboring source(s). 
Background counts were estimated from neighbouring source-free circular regions \footnote{The X-ray photons from the host galaxy are contained within 5 arcsecond from the centroid of NGC4993. No other X-ray sources lie in the close vicinity of GW170817. Therefore, in order to obtain the background-corrected photon counts for GW170817, we considered a  $\sim$$35''$ diameter source free circular region,  $\sim$$1'$ away from GW170817 (as per standard practice in X-ray image analysis).} having radii of at least $15''$. 
The PSF-corrected count rates were calculated using the \textit{srcflux} task, where we used \textit{psfmethod=arfcorr} to incorporate the energy-dependent PSF correction\footnote{The \textit{arfcorr} task can also be run independently, before \textit{srcflux}, to account for the photon energy dependence, but we found that this approach requires substantially larger computational time (4-5 hours) compared to the case where the \textit{arfcorr} parameter is set within \textit{srcflux} (few seconds of computational time required for the same energy step size). Nevertheless, the results from both these approaches was found to be the same.} as recommended by \cite{troja2022,Makhathini2021}, which follows the expression given below \citep[see also ][]{troja2022}.
\begin{equation*}
    r_s = \eta \left(N - \frac{B \cdot A_s}{A_b}\right) \Delta t^{-1}
\end{equation*}

\noindent Here, \( N \) and \( B \) represent the observed total and background counts within extraction portions of areas \( A_s \) and \( A_b \), respectively. 
The correction factors (which we found to lie between $\sim$0.7--1), accounted for by \textit{arfcorr} and based on energy-dependent aperture (PSF fraction), are represented by \( \eta \). \( \Delta t \) represents the exposure duration. 
% This is required to calculate the fraction of counts from the source that are located in the area at each energy. . 
% This formulation of applying the PSF corrected Count Rates is used by \textit{arfcorr} and \textit{srcflux} tasks. 

In our analysis we utilized the \textit{srcflux}\footnote{The \textit{bands} parameter in the \textit{srcflux} task, which takes the effective energy as the input, was calculated using the \textit{find\_mono\_energy} task with \textit{mean} and \textit{max} parameters and also independently using the \textit{dmtcalc} task within the energy range 0.3--10 keV. The effective energies and flux values calculated using these different approaches show a maximum variation of 2--7\% across all epochs. For our final analysis, we used the effective energies given by \textit{dmtcalc}, which vary between 3.3--3.8 keV across epochs 1--14.} task to compute count rates and fluxes (in the 0.3--10 keV band) of GW170817, along with their associated 1$\sigma$ uncertainties for each exposure/epoch. 
Guided by \cite{troja2022}, we also inspected the Energy Conversion Factor (ECF) values for each exposure. 
ECF measures the efficiency of the entire hardware and software system in converting the incoming X-ray counts into flux. 
We calculated ECF as per the following equation.
\[
\text{Flux (erg cm$^{-2}$ s$^{-1}$)} = \text{ECF} \times \text{Count Rate (counts s$^{-1}$)}
\]

\noindent the resultant ECF versus time plot is shown in Figure~\ref{fig:ECF}.
For single exposures that had no photon detections we found the 3$\sigma$ upper limits on the count rates and fluxes using the standard approach recommended\footnote{https://cxc.cfa.harvard.edu/ciao/threads/upperlimit/} with the \textit{srcflux} task.
The ECF values for such exposures is found using the above expression, but instead using just the upper limits on count rates and fluxes. 

The measured flux values (or upper limits) were converted into flux densities at 1 keV using a spectral index found iteratively by fitting the X-ray data points together with the other panchromatic afterglow data with a broken power law model (described in Sec.~\ref{sec:panchromatic}). 
The resultant convergence was on a spectral index of $\beta = -0.580\pm0.002$. 
Using this spectral index, 0.3--10 keV X-ray flux can be converted into a flux density, in units of Jy at 1 keV, through a multiplicative factor of 86027. Table~\ref{tab:flux_density} gives the resultant PSF-corrected count rates, fluxes and flux density values (at\,1\,keV\,$=2.4 \times 10^{17}$\,Hz) for each epoch, and Table~\ref{tab:Individual_obs} gives the PSF-corrected count rates in different energy bands along with the ECF and Flux values for each individual exposure (observation). 

In Figure~\ref{fig:lc-2methods} we present a comparison between the flux densities obtained from our reanalysis and those published previously by various research groups \citep{Makhathini2021,troja2022,hajela2022}. In this figure the errorbars on the \cite{hajela2022} points are based on uncertainties in the fluxes only; we have neglected the uncertainties that the authors indicated on the photon index ($\Gamma = 1 - \beta$).
The \cite{troja2022} data points correspond to their fluxes derived for a fixed value of $\Gamma=1.585$.
Generally we find that, between the techniques used by different groups, the initial upper limit varies by up to 200\%, while the individual flux density values vary by 50\% on an average although they generally all agree within their 1$\sigma$ uncertainties.

% \muskan{Describe ECF and how you calculated it. Also reference the ECF plot}. \myadav{added}
% \muskan{mention about srcflux here first! And write the following expression as an equation} \myadav{Done} 
% \muskan{explain what energy dependent aperture is -- note that the methodology described in this section should be accessible to even new users of CIAO/X-ray data processing} \myadav{Done}
% With the exception of the initial observation at 2.3 days (ObsID 18955), where the importance of the search is above 3$\sigma$, X-ray emission is constantly observed from the position of GW170817. 

To investigate the possible evolution of the X-ray spectrum at late times, which we first generated X-ray spectra for each exposure using the \textit{srcflux} task (same parameters as those used to calculate fluxes; described above). 
We then combined the X-ray spectra from various epochs and groups of epochs using the \textit{combine\_spectra} task. 
Using \textit{PyXspec} and following previous works \citep[e.g.][]{Makhathini2021,troja2022,hajela2020}, we modeled the spectra with the \textit{tbabs * powerlaw} model\footnote{The \textit{ztbabs} component was excluded, as it has been confirmed previously \citep{Makhathini2021,hajela2020}, as well as in our analysis, that there is no contribution from intrinsic galactic absorption.}, where \textit{tbabs} accounts for the neutral absorbing column attributed to our own Galaxy, fixed at \( 1.1 \times 10^{21} \, \text{cm}^{-2} \) \citep[also see][]{Willingale2013,troja2019}. 
We found that the $\beta = -0.58$ is consistent with all spectra well within 1$\sigma$, hence we fixed $\beta$ to this value and calculated the reduced $\chi^2$ for each set of spectra (Figure~\ref{fig:xray-spec}). 
The spectra together and their power-law models and corresponding $\chi^2_{\nu}$ values are shown in Figure~\ref{fig:xray-spec}.

\begin{figure*}
    \centering
    \includegraphics[width=0.67\textwidth]{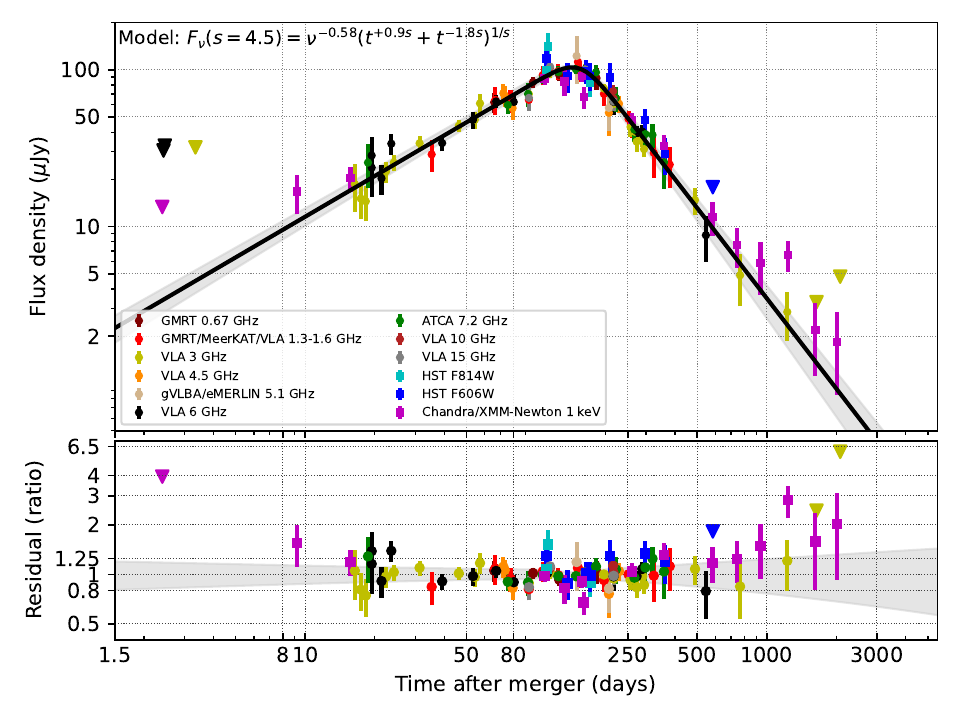}
    \caption{Upper panel: Broken power-law fit (black curve) to the afterglow light curve, color coded according to the observing frequency, (all data points have 1$\sigma$ errorbars; upper limits are 3$\sigma$) using the uniform dataset presented in this work. The light curve is scaled to 3 GHz using the best-fit spectral index ($\nu^{-0.580}$) derived from the MCMC power-law fitting. Lower panel: The residual of the broken power-law fit.}
    \label{fig:lc}
\end{figure*}

\begin{figure*}  
    \centering
    \includegraphics[width=0.62\textwidth]{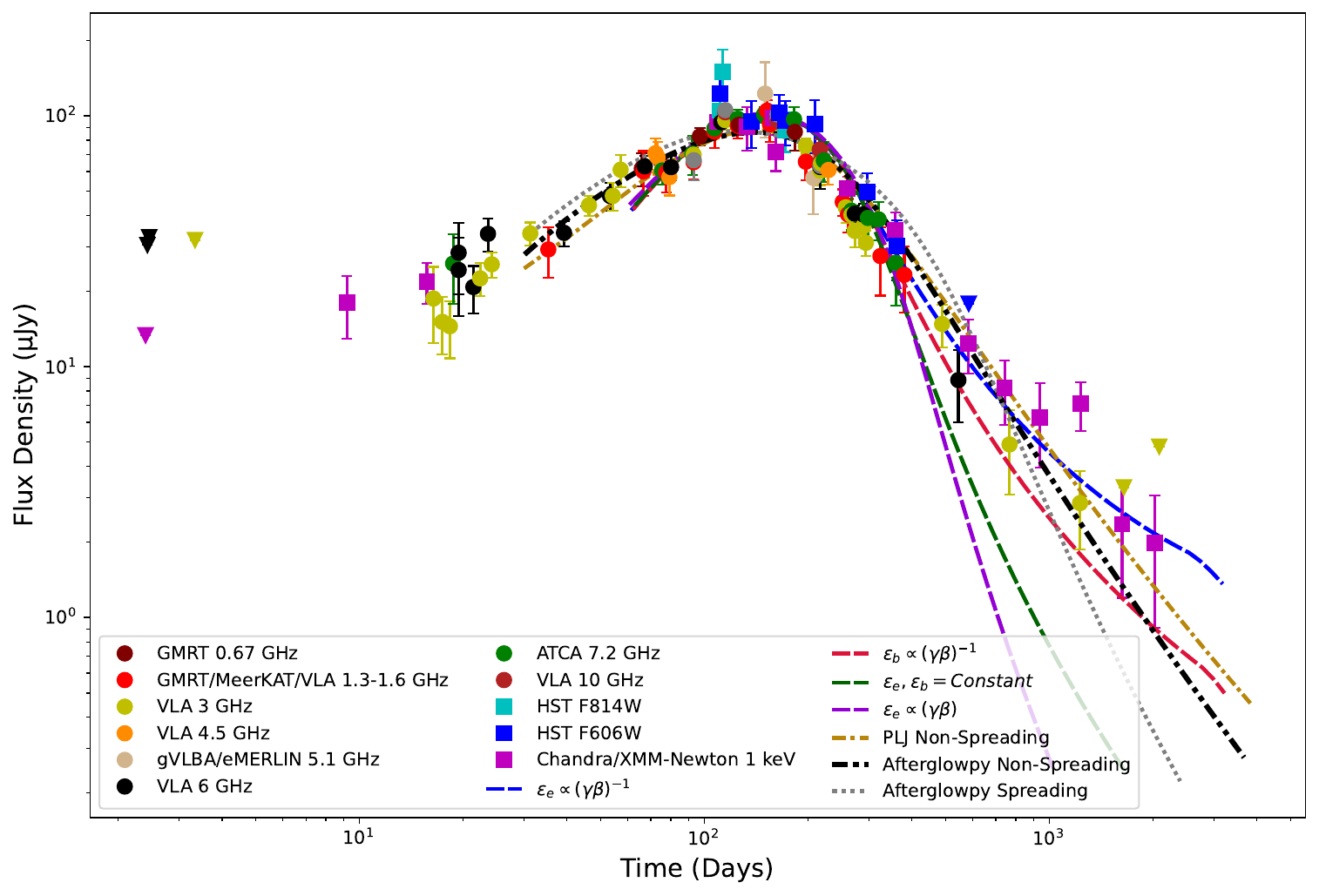}
    \caption{Hydrodynamical models plotted with the panchromatic afterglow data for GW170817. The color coding of the data points is the same as in Figure~\ref{fig:lc}. The dashed curves denote models for which one of the microphysical parameters $\epsilon_e$ or $\epsilon_B$ is a function of the ejecta speed \citep{Govreen-Segal2024}: a) $\propto (\gamma\beta)^{-1}$ (blue and red dashed curves), b) $\propto (\gamma\beta)$ (magenta), and c) constant (green). The non-spreading jet models from afterglowpy (black) and this work (brown) are shown as dot-dashed curves. The dotted grey curve represents a spreading jet model from afterglowpy. The non-spreading jet models and the $\epsilon_e \propto (\gamma\beta)^{-1}$ jet model are most consistent with the observational data.}
    \label{fig:HSD}
\end{figure*}

\begin{figure}
    \centering
    \includegraphics[width=0.5\textwidth]{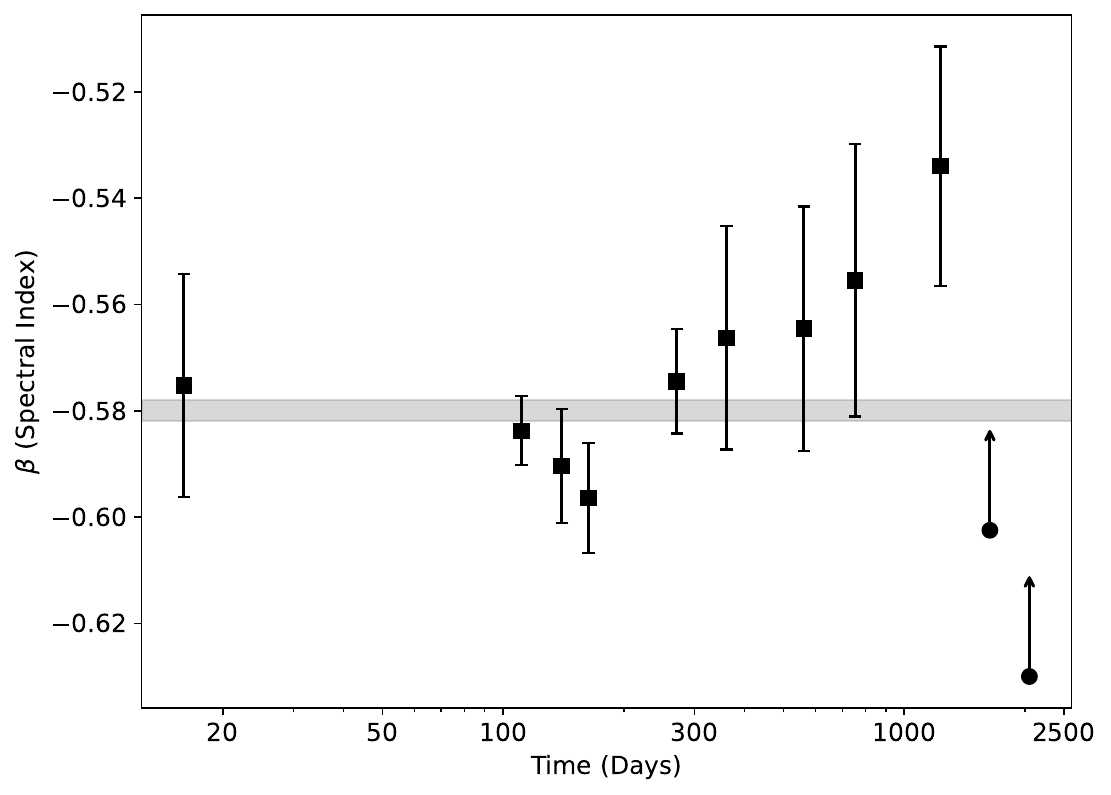}
    \caption{Spectral index (\(\beta\)) values derived from radio, optical, and X-ray data at various times for GW170817. Two lower limits at later times, derived from radio upper limits, are consistent with \(\beta = -0.580\pm0.002\) (shaded region), as obtained from the broken power-law fit. This consistency suggests that a single power-law index suffices to describe the entire multiwavelength dataset.
}
    \label{fig:Spectral_index}
\end{figure}

\begin{figure}
    \centering
    \includegraphics[width=0.5\textwidth]{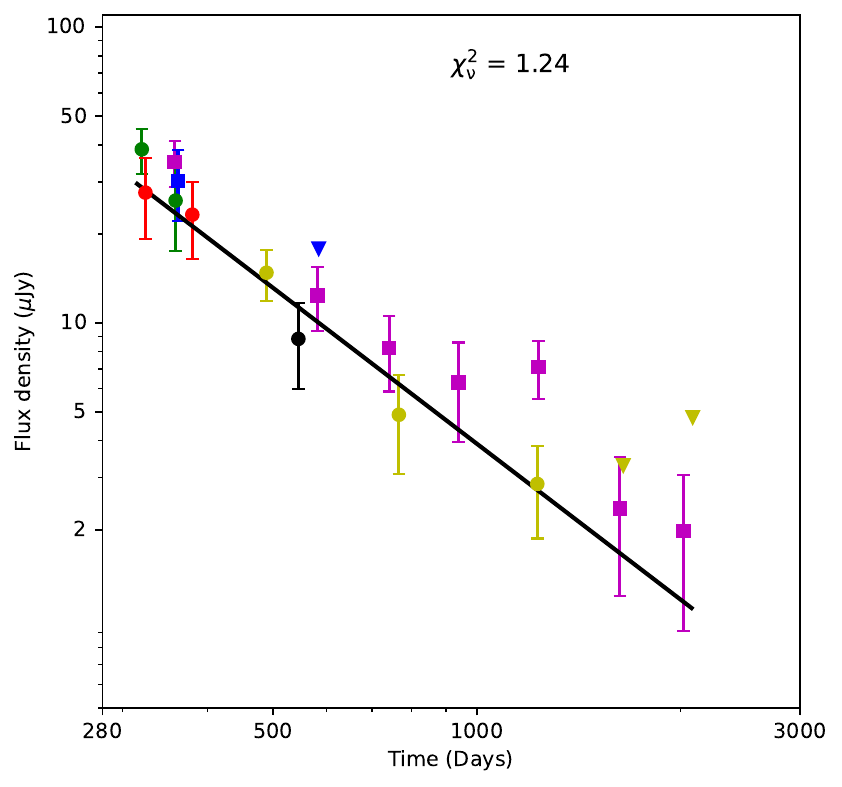}
    \caption{Afterglow data of GW170817 (post-peak decline) fitted with the power-law \( t^{-1.758} \), as found from the broken power-law fit (Sec.~\ref{sec:panchromatic}) The fit yields a reduced chi-square value of 1.24, indicating that a single power-law component fits the decline part well, without the need to invoke an additional afterglow component to explain the late-time afterglow emission.}
    \label{fig:late_time_fit}
\end{figure}

\section{The Panchromatic Afterglow Light Curve And Model Fitting}\label{sec:panchromatic}

% \muskan{describe the methodology for gathering the most up to date multiwavelength dataset, the fitting procedure and reference all remaining figures.}
We obtained the panchromatic dataset from \cite{Makhathini2021}, revised the HST data with the reprocessed dataset from  \cite{Kilpatrick2022} and the Chandra X-ray dataset from this work. This revised and updated panchromatic afterglow dataset of GW170817 is available on GitHub\footnote{\url{https://github.com/kmooley/GW170817/}}. Aligned with previous afterglow studies \citep{Beuermann1999,alexander2018,mooley2018-strongjet,troja2019,Makhathini2021}, we used a smoothly broken power-law model to perform a MCMC fit the panchromatic afterglow light curve of GW170817:

\[
F_{\nu}(t) = F_p \left( \frac{\nu}{\nu_p} \right)^{\beta} \left[ \frac{1}{2} \left( \frac{t}{t_p} \right)^{-s\alpha_1} + \frac{1}{2} \left( \frac{t}{t_p} \right)^{-s\alpha_2} \right]^{-1/s},
\]

\noindent where \(F_{\nu}\) is the flux density at the observing frequency \(\nu\), and \(t\) is the time since the merger. 
% \muskan{please add a sentence here saying: "We obtain best-fit values of  .., .., .., .., .., ... for the parameters of the smoothly broken power law model (68\% confidence interval"}.
We obtained the best-fit values for the parameters of the smoothly broken power-law model, with 68\% confidence errorbars, as follows: \(F_p = 102.2^{+3.44}_{-3.07} \, \mu\mathrm{Jy}\)
, \(t_p = 150.5^{+3.27}_{-3.02} \, \mathrm{days}\), \(\alpha_1 = 0.85^{+0.03}_{-0.03} \), \(\alpha_2 = -1.758^{+0.07}_{-0.08} \), \(\beta = -0.58^{+0.002}_{-0.002} \), and \(s = 4.57^{+1.97}_{-1.06} \). 
% Here, \(F_p\) represents the peak flux density, and \(t_p\) is the time at which the light curve peaks. The parameters \(\alpha_1\) and \(\alpha_2\) denote the power-law rise and decay slopes, respectively, while \(\beta\) represents the spectral index, and \(s\) is the smoothness parameter of the transition.
The parameter \(F_p\) represents the peak flux density at \(\nu_p = 3 \, \mathrm{GHz}\), while \(t_p\) is the time at which the light curve peaks. The smoothness of the transition between the rising and decaying phases of the light curve is governed by the parameter \(s\). The spectral index \(\beta\) defines the frequency dependence, and \(\alpha_1\) and \(\alpha_2\) are the power-law indices for the rise and decay of the light curve, respectively. 
The panchromatic afterglow data, normalized to 3 GHz (where there afterglow is sampled most regularly) and fit with the above model, are shown in Figure~\ref{fig:lc} along with the residuals. 
The corresponding corner plot is shown in Appendix A (Figure~\ref{fig:corner_plot}).

To understand the implications of the late-time afterglow on jet dynamics (discussed in Sec.~\ref{sec:panchromatic}), we compared the late-time (250--2020 days post-merger) panchromatic light curve with various jet models.
This comparison is shown in Figure~\ref{fig:HSD}.
We fit two non-spreading jet models (power-law/gaussian) from \cite{hotokezaka2019}  ($\chi^2_{\nu}$ = 1.4, for time$>$300 days post merger) and \texttt{afterglowpy} \citep{ryan2020} ($\chi^2_{\nu}$ = 1.4) to the panchromatic data.
We also fit a spreading Gaussian jet model using \texttt{afterglowpy} ($\chi^2_{\nu}$ = 4.0).
Model light curves from hydrodynamical simulations carried out by \cite{Govreen-Segal2024}, incorporating microphysical parameters ($\epsilon_e$ or $\epsilon_B$)(for $\epsilon_e \propto (\gamma \beta)^{-1}$ the $\chi^2_{\nu}$ = 0.7; for $\epsilon_b \propto (\gamma \beta)^{-1}$ the $\chi^2_{\nu}$ = 1.8; for  $\epsilon_e \propto (\gamma \beta)$ the $\chi^2_{\nu}$ = 5.7 and for constant $\epsilon_e,\epsilon_b$ the $\chi^2_{\nu}$ = 4.8) varying with jet speed and for unchanging microphysical parameters, aligned with the peak of the afterglow light curve are also shown in Figure~\ref{fig:HSD}.

     To investigate whether there is any temporal evolution of the spectral index (\(\beta\)), we performed power-law fits to the panchromatic afterglow data (radio, optical, and X-ray) at various epochs, wherever contemporaneous observations were available at two or more wavelengths\footnote{During the rise phase of the light curve (at 16 days) and the declining phases (562, 754, 1231 days), we used the available  radio and X-ray data to calculate the spectral index. Around the peak of the light curve (111, 139, 163, 271, 360 days), data from radio, optical, and X-ray observations were incorporated into the calculation of (\(\beta\)). For the late-time observations (1635, 2051 days), radio upper limits, along with X-ray data, were used to estimate the lower limits of the spectral index.}. 
The resulting values of \(\beta\) are plotted in Figure~\ref{fig:Spectral_index}. 

% \muskan{Note for AMan and Muskan: please reword this section properly to make it publication-quality and fill in all the missing references.}

\section{Discussion}\label{sec:discussion}
We now bring all of the panchromatic data and previous analyses into context to investigate the implications of the late-time afterglow ($>$5 years post-merger) of GW170817.

\begin{enumerate}
\item \textbf{Shallow post-break slope:}
    We model the afterglow as a broken power-law in both frequency and time, given by $F_\nu \propto \nu^{\beta} t^{-\alpha}$, where $\alpha = 1.76\pm0.08$ and $\beta = -0.580\pm0.002$ during the light curve decline. 
    The spectral index $\beta$ is related to the power-law index of the electron energy distribution, $p$, as $\beta = -(p-1)/2$, which indicates $p\approx2.2$.
     \cite{sari1999} analytically show that the monochromatic afterglow flux after the jet break asymptotically declines as $F_{\nu} \propto t^{-p}$ assuming the jet spreads exponentially with time.  
    Relativistic hydrodynamic simulations of jets suggest a steeper post-break decline rate, approximately $t^{-2.6}$
    for $p=2.2$ \citep[e.g.][]{vaneerten2013,govreen2023analytic}.
    However, the decline rate of the observed afterglow of GW170817, $t^{-1.8}$, is significantly shallower than these predictions.
    In Figure \ref{fig:HSD} we show a visual comparison between the different models fit to/overlayed on the afterglow light curve of GW170817. 
    We find that the non-spreading jet models and the models where microphysical parameters ($\epsilon_e$ and $\epsilon_B$) are varying with jet speed, are more consistent with the observed $t^{-1.8}$ decline while the hydrodynamical models assuming constant microphysical parameters, and numerical models incorporating spreading, predict a substantially steeper afterglow decline that is not seen in GW170817 (see also Sec.~\ref{sec:panchromatic}).
    This inconsistency may suggest either additional physical mechanisms affecting the emission such as an additional energy source \citep[e.g.,][]{ryan2024,Sadeh2024} and/or the velocity dependence of the efficiencies of electron acceleration and magnetic-field amplification \citep[see, e.g.,][ and Figure \ref{fig:HSD}]{govreen2023analytic}.
    %or potential inaccuracies in the assumed power-law index.
    %\quad \text{(homogeneous medium)}, or, 
%\[
%p = \frac{10\alpha + 21}{15} \quad \text{(homogeneous medium)}
%\]
%or 
%\[
%p = \frac{6\alpha + 5}{7} \quad \text{(wind medium)}.
%\]

%Using these expressions, we obtain \( p \approx 2.16 \) and \( p \approx 2.56 \) for the homogeneous medium, and \( p \approx 2.21 \) for the wind medium. Whereas assuming that the circumburst medium density follows a power-law with radius \( r \) with index \( k \), i.e., proportional to \( r^{-k} \), the value of \( k \) and its uncertainty can be calculated directly from \(\alpha\) and \(\beta\) as follows:

%\[
%k = \frac{5\alpha - 15\beta + 3}{\alpha - 4\beta + 2}.
%\]

%Using our fitted values, we obtain \( k = 2.14 \), which is consistent with a wind-like medium (\(k = 2\)) rather than a homogeneous medium (\(k = 0\)).

    \item \textbf{Constant spectral index across time:}
    In Figure~\ref{fig:xray-spec} we show the X-ray spectrum across 2400 days post-merger. 
    There is no significant evolution as evident from the $\chi^2_{\nu}$ values of the $\beta = -0.58$ power law fits to the different epochs. 
    In Figure~\ref{fig:Spectral_index} we show the spectral index as a function of time using contemporaneous observations of the afterglow across radio-optical-X-ray wavelengths. 
    Although a hint of a 2$\sigma$ deviation in spectral index from the mean of $-0.58$ are seen around 1200 days post-merger, we do not find substantial evidence for this from Figure~\ref{fig:xray-spec}.
    This point has been discussed at length by \cite{hajela2022,troja2022} and we suggest that this data point, considered within the context of $>$100 data points of the panchromatic light curve, is statistically not a significant outlier.
    In order to further verify whether there is any evolution in the spectral index, we fit two other models, a linearly varying spectral index and linear variation in the spectral index with a break, to the spectral index verses time data. Our F-tests \citep{Bevington1992} performed on these models suggest that a constant spectral index provides a statistically robust fit compared to the other two models.
    The constancy of the the spectral index across time then suggests that the power-law index of accelerated electrons, $p$, does not significantly change over time. Note that
    \cite{Keshet2005PhRvL} predicts that the value of $p$ is $\approx 2.2$ in the ultra-relativistic limit. 
    The power-law index decreases to the non-relativistic value, $p=2$, in the velocity range of $1\lesssim \gamma \beta \lesssim 10$, corresponding to the spectral index changing from $-0.55$ to $-0.5$ \citep[see][in the context of GW170817]{Takahashi2022MNRAS}. 
    The observed values shown in Figure \ref{fig:Spectral_index} are marginally consistent with such evolution.
    
    %\muskenta{Do you think the spectral index slightly increases with time? This change in $p$ predicts an evolution from $-0.55$ to $-0.5$. Although we cannot say that, the data seem consistent with Keshet and Waxman.}
    %This could imply that the observed radiation is dominated by a single mechanism, such as synchrotron emission, without substantial variation in external influences, such as density changes or magnetic field fluctuations.
    % Figure~\ref{fig:xray-spec}, \ref{fig:Spectral_index}, \ref{fig:late_time_fit}

    \item \textbf{Synchrotron Cooling Limit and X-ray Band:}
    From Figure~\ref{fig:xray-spec} we find that a power law with $\beta = -0.58$ gives good fits to the spectra at all epochs ($\chi_\nu^2\simeq1$), even for the two epochs observed beyond 2000 days post-merger. 
    % The lack of any evolution in the spectral index, at especially the later epochs, indicates that the cooling break has likely not passed through the X-ray band $\sim$4--6 years post-merger and that a single component (jet afterglow) satisfactorily explains the late-time afterglow without the need for invoking alternate emission mechanisms (e.g. afterglow from the kilonova ejecta). 
    We therefore conclude that the synchrotron cooling break has not yet extended into the X-ray band. 
    The cooling break in the mildly relativistic regime ($\gamma\beta\approx 1$) is expected to occur at the cooling frequency \citep{sari1999}:
% \[
% \nu_c \approx 10^{17} \, \text{Hz} \left( \frac{\gamma}{2} \right)^{-4} \left( \frac{n}{10^{-4} \, \text{cm}^{-3}} \right)^{-3/2} \left( \frac{\epsilon_B}{10^{-2}} \right)^{-3/2}  
%      \left( \frac{t}{2500 \, \text{days}} \right)^{-2},
% \]  
\begin{align*}
\nu_c &\approx 5\cdot 10^{17} \, \text{Hz} (\gamma\beta)^{-3} 
\left( \frac{n}{10^{-4} \, \text{cm}^{-3}} \right)^{-3/2} 
\left( \frac{\epsilon_B}{10^{-2}} \right)^{-3/2} \notag \\
&\hspace{4.5cm} \left( \frac{t}{2400 \, \text{days}} \right)^{-2},
\end{align*}

    where $\gamma$ is the Lorentz factor of the outflow,
    $n$ is the number density of the ISM, $\epsilon_B$ is the efficiency of magnetic field amplification, and $t$ is the observer time.
    The observed spectra of the late-time afterglow is described well with a single power-law from the radio to X-ray, indicating the cooling frequency is higher than $\sim 10\, {\rm keV}$. Assuming the outflow emitting the late-time afterglow is mildly relativistic, this constraint implies\\
\[
\epsilon_B n \lesssim 3\cdot 10^{-7}\,{\rm cm^{-3}}(\gamma\beta)^{-2}.
\]   
    This is consistent with the values derived from the light curve and centroid motion modelings \citep{Takahashi2021MNRAS,ryan2024,Sadeh2024}.

    \item \textbf{Lack of evidence for a new afterglow component:}
    In Figure~\ref{fig:late_time_fit} we show that the decline part of the panchromatic afterglow light curve is fit with a single power-law model \( t^{-1.758} \) (obtained from the smoothly broken power-law fit described in Sec.~\ref{sec:panchromatic}) to give $\chi_\nu^2 \simeq 1$. We also fit a broken power-law model to the decline part of the lightcurve and the corresponding F-test \citep{Bevington1992} comparing both these models indicates a $<$10\% variance reduction, with the broken power-law model compared to the single power-law model.
    Thus, when considered together with radio and optical data, the X-ray data (including late-time emission) can be satisfactorily explained using just a single component. 
    The constant spectral index across time (discussed above) fortifies this argument. 
    Stated differently, no new afterglow component has thus emerged, suggesting the afterglow is consistent with a single emission process without significant contributions from additional ejecta or refreshed shocks. 
    This is in accordance with \cite{ryan2024} who recently found that the evidence for a new afterglow component in the X-rays is $\lesssim 2\sigma$.
    The fact that the late-time afterglow of GW170817 is consistent with the same power-law decline suggests that the jet must be still in the mildly relativistic, $\gamma\sim2$, phase $\sim$2000 days post-merger.

    % \item \textbf{If \( t^{-p} \) does not precisely hold for GRBs.:}
    % In the scenario where the observing frequency is between the synchrotron cooling and characteristic frequencies ($\nu_m<\nu_X<\nu_c$), the value of the electron power law index ($p$) can be obtained in two ways: $p=1-2\beta$ and $p=1+4\alpha/3$ \citep[$F_\nu \propto \nu^{\beta} t^{-\alpha}$;][]{granot_sari2002}. 
    % This method has been traditionally used to infer the value $p$ for populations of GRBs \citep[e.g.][]{fong2015}.
    % If this method is used for GW170817 then we get $p=2.2$ (as stated previously) and $p=3.3$ respectively.
    % % We will need to revisit the conclusions drawn from previous short GRBs or GRBs in general. The assumption of post-break \( t^{-p} \) may not be robust.
    % \muskan{Note to Kenta: We need to show that parameters like energy etc, derived using $p$ calculated as above, may give discrepant results. Could you please help complete this point? If we can add something slightly quantitative that will be excellent.}

\end{enumerate}

\section*{Acknowledgements}
We are grateful to Muskan Yadav for assistance with X-ray data processing and data fitting. We extend our gratitude to the Chandra X-ray Center (CXC) staff for their invaluable assistance in resolving the challenges encountered during data analysis.
K.H. is supported by the JST FOREST Program (JPMJFR2136) and the JSPS Grant-in-Aid for Scientific Research (20H05639, 20H00158, 23H01169, 23H04900). We acknowledge support from the Anusandhan National Research Foundation (ANRF), Government of India, through the Prime Minister Early Career Research Grant (ANRF/ECRG/2024/005949/PMS).

\section*{Data Availability}
The revised and updated panchromatic afterglow dataset of GW170817
is available on \url{https://github.com/kmooley/GW170817/}.

\newpage
\begin{table*}
    \centering
    \footnotesize
    \caption{X-ray Afterglow Measurements of GW170817}
    \label{tab:flux_density}
    % {\centering \caption*{\textbf{X-ray Afterglow Measurements of GW170817 \\[1em]}}}
    % \caption{X-ray Afterglow Measurements of GW170817 \\[1em]}
    % \vspace{5pt}
    % \large
    % \raggedright
    % \renewcommand{\arraystretch}{1.1}
    \begin{tabular}{cccccccc}
        \hline
        \hline
        \multicolumn{1}{c}{Epoch} & \multicolumn{1}{c}{T-T0} & \multicolumn{1}{c}{Exposure} & \multicolumn{1}{c}{Count $\mathrm{rate}^{a}$} & \multicolumn{1}{c}{Flux $\mathrm{Density}^{b}$} & \multicolumn{1}{c}{$\mathrm{Flux}^{c}$} & \multicolumn{1}{c}{ObsID} \\
        \multicolumn{1}{c}{} & \multicolumn{1}{c}{(d)} & \multicolumn{1}{c}{(ks)} & \multicolumn{1}{c}{[ 0.5 - 7 keV ]} & \multicolumn{1}{c}{[ 1keV ]} & \multicolumn{1}{c}{[ 0.3 - 10 keV ]} & \multicolumn{1}{c}{} \\
        \hline
         1 & 2.33 & 24.6 & $<$3.28 & $<$4.68 & $<$5.44 & 18955 \\
                &  &  &  &  &  &  \\
         2 & 9.2 & 49.4 & $3.01^{+0.94}_{-0.78}$ & $4.34^{+1.35}_{-1.12}$ & $5.05^{+1.57}_{-1.31}$ & 19294 \\
                &  &  &  &  &  &  \\
         3 & 15.7 & 93.4 & $3.67^{+0.7}_{-0.6}$ & $5.28^{+1.03}_{-0.86}$ & $6.14^{+1.2}_{-1.0}$ & 18988, 20728 \\
                &  &  &  &  &  &  \\
         4 & 108 & 98.8 & $15.7^{+1.3}_{-1.4}$ & $22.88^{+1.98}_{-1.98}$ & $26.6^{+2.3}_{-2.3}$ & 20860, 20861 \\
                &  &  &  &  &  &  \\
         5 & 158 & 104.9 & $16.1^{+1.4}_{-1.3}$ & $23.65^{+1.98}_{-1.89}$ & $27.5^{+2.3}_{-2.2}$ & 20936, 20937, 20938 \\
                &  &  &  &  &  & 20939, 20945 \\
                &  &  &  &  &  &  \\
         6 & 260 & 96.8 & $8.28^{+0.99}_{-0.97}$ & $12.39^{+1.46}_{-1.46}$ & $14.4^{+1.7}_{-1.7}$ & 21080, 21090 \\
                &  &  &  &  &  &  \\
         7 & 358 & 67.2 & $5.61^{+1.03}_{-0.93}$ & $8.46^{+1.60}_{-1.41}$ & $9.84^{+1.86}_{-1.64}$ & 21371 \\
                &  &  &  &  &  &  \\
         8 & 582 & 98.3 & $1.92^{+0.5}_{-0.4}$ & $2.99^{+0.77}_{-0.69}$ & $3.48^{+0.9}_{-0.8}$ & 21322, 22157, 22158 \\
                &  &  &  &  &  &  \\
         9 & 741 & 98.9 & $1.23^{+0.4}_{-0.3}$ & $1.99^{+0.69}_{-0.52}$ & $2.31^{+0.8}_{-0.6}$ & 21372, 22736, 22737 \\
                &  &  &  &  &  &  \\
         10 & 940 & 96.6 & $0.91^{+0.38}_{-0.3}$ & $1.52^{+0.60}_{-0.52}$ & $1.77^{+0.7}_{-0.6}$ & 21323, 23183, 23184 \\
                &  &  &  &  &  & 23185 \\
                &  &  &  &  &  &  \\
         11 & 1234 & 189 & $0.98^{+0.27}_{-0.23}$ & $1.72^{+0.43}_{-0.34}$ & $2.0^{+0.5}_{-0.4}$ & 22677, 24887, 24888 \\
                &  &  &  &  &  & 24889, 23870,24923, \\
                &  &  &  &  &  & 24924 \\
                &  &  &  &  &  &  \\
         12 & 1625 & 193.6 & $0.31^{+0.17}_{-0.13}$ & $0.57^{+0.32}_{-0.24}$ & $0.66^{+0.37}_{-0.28}$ & 23869, 24336, 24337 \\
                &  &  &  &  &  & 25527, 25733,25734 \\
                &  &  &  &  &  & 26223 \\
                &  &  &  &  &  &  \\
         13 & 2020 & 182.8 & $0.25^{+0.16}_{-0.12}$ & $0.48^{+0.31}_{-0.22}$ & $0.557^{+0.36}_{-0.26}$ & 25528, 27088, 27089 \\
                &  &  &  &  &  & 27090, 27091,27731 \\
                &  &  &  &  &  &  27752, 27753, 27754 \\
                &  &  &  &  &  &  \\
         % 14 & 2440 & 172 & $0.28^{+0.17}_{-0.13}$ & $0.57^{+0.34}_{-0.26}$ & $0.66^{+0.4}_{-0.3}$ & 28358, 28523, 28525 \\
         %        &  &  &  &  &  & 28526, 28527,29370 \\
         %        &  &  &  &  &  &  29377, 29395, 29397 \\
         %        &  &  &  &  &  &  29398, 29399 \\
              
        \hline
    \end{tabular}
   \vspace{0.5em}
   {\centering \\ $^{a}$ Count rates are in units of $10^{-4}$ cts $s^{-1}$. All the values are corrected for PSF losses. \\
     $^{b}$ Flux Density in units of $10^{-4}$ \textmu{}Jy. All data points have 1$\sigma$ errorbars; upper limits are 3$\sigma$. \\
     $^{c}$ Fluxes in units of $10^{-15}$ erg $cm^{-2}$ $s^{-1}$. Values are corrected for Galactic extinction.\\}
    % \raggedright}
\end{table*}

%%%%%%%%%%%%%%%%%%%% REFERENCES %%%%%%%%%%%%%%%%%%

% The best way to enter references is to use BibTeX:
\newpage
\bibliographystyle{mnras}
\bibliography{example} % if your bibtex file is called example.bib

% Alternatively you could enter them by hand, like this:
% This method is tedious and prone to error if you have lots of references
%\begin{thebibliography}{99}
%\bibitem[\protect\citeauthoryear{Author}{2012}]{Author2012}
%Author A.~N., 2013, Journal of Improbable Astronomy, 1, 1
%\bibitem[\protect\citeauthoryear{Others}{2013}]{Others2013}
%Others S., 2012, Journal of Interesting Stuff, 17, 198
%\end{thebibliography}

%%%%%%%%%%%%%%%%%%%%%%%%%%%%%%%%%%%%%%%%%%%%%%%%%%

%%%%%%%%%%%%%%%%% APPENDICES %%%%%%%%%%%%%%%%%%%%%

% \appendix

\newpage
\section*{Appendix}

\begin{figure*}
    \centering
    \includegraphics[width=\textwidth]{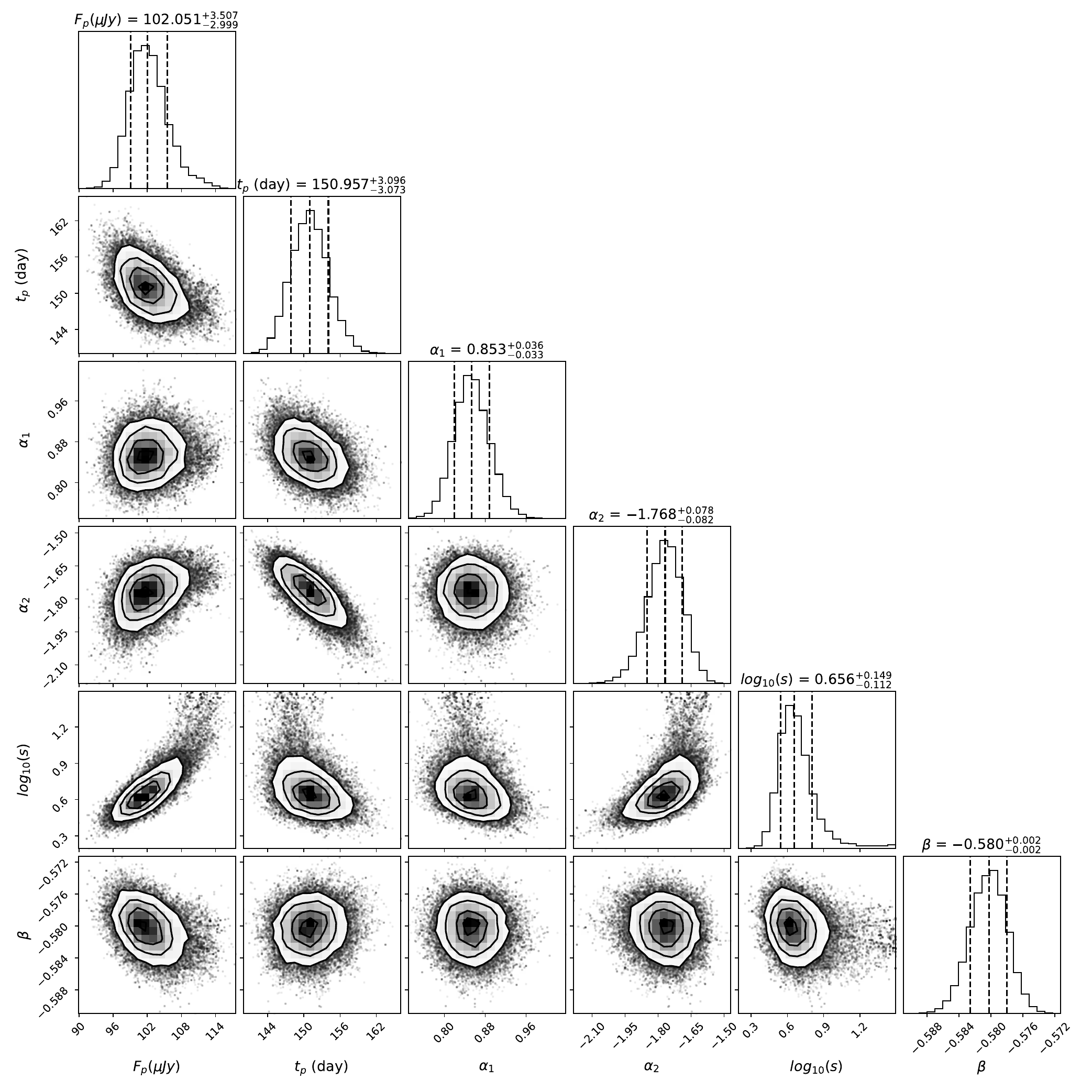}
    \caption{Corner plot for broken power-law fit to the light curve presented in Figure 1. Here, $\beta$ is the spectral index, $F_p$ is the flux density at 3 GHz at the light curve peak, $t_p$ is the light curve peak time, and $\alpha_1$ and $\alpha_2$ are the power-law rise and decay slopes, respectively.}
    \label{fig:corner_plot}
\end{figure*}

\newpage
\begin{Large}
\renewcommand{\arraystretch}{1.35}
\onecolumn
\begin{longtable}{|c|c|c|c|c|c|c|}
    \caption{Individual observations with count rate and fluxes} \label{tab:Individual_obs} \\  
    \hline
    \multicolumn{1}{|c|}{ObsID} & \multicolumn{1}{c|}{T-T0 } & \multicolumn{1}{c|}{Exposure } & \multicolumn{1}{c|}{Count $\mathrm{rate}^{a}$} & \multicolumn{1}{c|}{ECF} & \multicolumn{1}{c|}{$\mathrm{Flux}^{b}$} & \multicolumn{1}{c|}{PI}\\
    \multicolumn{1}{|c|}{} & \multicolumn{1}{c|}{(d)} & \multicolumn{1}{c|}{(ks)} & \multicolumn{1}{c|}{[0.5 - 7 keV]} & \multicolumn{1}{c|}{} & \multicolumn{1}{c|}{[0.3 - 10 keV]} & \multicolumn{1}{c|}{}  \\
    \hline
    \endfirsthead
    \hline
    \multicolumn{7}{|c|}{\textbf{Continued from previous page}}  \\
    \hline
    \multicolumn{1}{|c|}{ObsID} & \multicolumn{1}{c|}{T-T0 } & \multicolumn{1}{c|}{Exposure } & \multicolumn{1}{c|}{Count $\mathrm{rate}^{a}$} & \multicolumn{1}{c|}{ECF} & \multicolumn{1}{c|}{$\mathrm{Flux}^{b}$} & \multicolumn{1}{c|}{PI}\\
    \multicolumn{1}{|c|}{} & \multicolumn{1}{c|}{(d)} & \multicolumn{1}{c|}{(ks)} & \multicolumn{1}{c|}{[0.5 - 7 keV]} & \multicolumn{1}{c|}{} & \multicolumn{1}{c|}{[0.3 - 10 keV]} & \multicolumn{1}{c|}{}  \\
    \hline
    \endhead
    \hline
        18955    & 2.3      & 24.95         & $<$3.28                &  1.65 &  $<$5.44               &   Fong              \\ \hline
        19294    & 9.2      & 50.06         & $3.01^{+0.94}_{-0.78}$ & 1.66 & $5.05^{+1.57}_{-1.31}$  &   Troja      \\ \hline
        20728    & 15.4     & 47.31         & $4.46^{+1.13}_{-0.98}$ & 1.66 & $7.44^{+1.9}_{-1.63}$   &   Haggard, Troja        \\ \hline
        18988    & 15.9     & 47.31         & $2.82^{+0.96}_{-0.78}$ & 1.66 & $4.73^{+1.61}_{-1.31}$  &   Wilkes     \\ \hline
        20860    & 108      & 74.09         & $15.9^{+1.60}_{-1.50}$   & 1.68 & $26.8^{+2.60}_{-2.60}$    &   Wilkes   \\ \hline
        20861    & 111.1    & 24.74         & $15.8^{+2.80}_{-2.60}$   & 1.67 & $26.7^{+4.90}_{-4.20}$    &   Wilkes     \\ \hline
        20936    & 153.6    & 67.16         & $19.5^{+2.60}_{-2.60}$   & 1.68 & $33.1^{+4.50}_{-4.40}$    &   Wilkes     \\ \hline
        20938    & 157.1    & 20.77         & $19.5^{+4.00}_{-3.40}$     & 1.69 & $33.3^{+6.70}_{-5.90}$    &   Wilkes     \\ \hline
        20938    & 157.1    & 20.77         & $19.5^{+4.00}_{-3.40}$     & 1.69 & $33.3^{+6.70}_{-5.90}$    &   Wilkes     \\ \hline
        20937    & 158.9    & 31.75         & $15.6^{+3.10}_{-2.80}$   & 1.69 & $26.6^{+5.20}_{-4.70}$    &   Wilkes      \\ \hline
        20939    & 159.9    & 15.86         & $11.9^{+2.70}_{-2.26}$  & 1.70 & $20.4^{+4.50}_{-4.00}$    &   Wilkes   \\ \hline
        20945    & 163.7    & 22.25         & $11.8^{+3.40}_{-2.80}$   & 1.69 & $20.2^{+5.70}_{-4.90}$    &   Wilkes    \\ \hline
        21080    & 259.2    & 14.22         & $7.78^{+1.38}_{-1.25}$ & 1.72 & $13.5^{+2.30}_{-2.20}$    &   Wilkes  \\ \hline
        21090    & 260.8    & 50.78         & $9.03^{+1.57}_{-1.39}$ & 1.72 & $15.7^{+2.70}_{-2.40}$    &   Wilkes    \\ \hline
        21371    & 358.6    & 46            & $5.61^{+1.03}_{-0.93}$ & 1.74 & $9.84^{+1.86}_{-1.64}$  &   Troja    \\ \hline
        21322    & 581      & 35.64         & $1.91^{+0.94}_{-0.70}$  & 1.80 & $3.49^{+1.70}_{-1.28}$   &   Margutti    \\ \hline
        22157    & 581.9    & 38.19         & $2.05^{+0.91}_{-0.71}$ & 1.80 & $3.72^{+1.66}_{-1.29}$  &   Margutti     \\ \hline
        22158    & 583.6    & 24.93         & $1.84^{+1.14}_{-0.81}$ & 1.81 & $3.35^{+2.08}_{-1.48}$  &   Margutti   \\ \hline
        21372    & 740      & 40            & $0.51^{+0.54}_{-0.33}$ & 1.85 & $0.97^{+1.01}_{-0.62}$ &  Troja       \\ \hline
        22736    & 742.2    & 33.61         & $1.33^{+0.84}_{-0.60}$ & 1.85 & $2.46^{+1.56}_{-1.11}$   &  Troja    \\ \hline
        22737    & 743.1    & 25.25         & $2.21^{+1.21}_{-0.89}$ & 1.85 & $4.12^{+2.24}_{-1.67}$  &   Troja     \\ \hline
        21323    & 935.5    & 24.29         & $1.38^{+1.03}_{-0.69}$ & 1.92 & $2.65^{+1.98}_{-1.32}$  &   Margutti   \\ \hline
        23183    & 939.1    & 16.28         & $<$4.97                & 1.91 & $<$9.52                 &   Margutti      \\ \hline
        23184    & 940.6    & 19.85         & $17^{+1.25}_{-0.84}$   & 1.92 & $3.3^{+2.40}_{-1.63}$    &   Margutti     \\ \hline
        23185    & 941.6    & 36.18         & $0.58^{+0.60}_{-0.36}$ & 1.92 & $1.14^{+1.17}_{-0.70}$   &   Margutti    \\ \hline
        22677    & 1209.7   & 29.69         & $0.70^{+0.70}_{-0.43}$ & 2.0 & $1.42^{+1.40}_{-0.88}$   &    Margutti  \\ \hline
        24887    & 1212.3   & 26.72         & $0.80^{+0.80}_{-0.49}$ & 2.0 & $1.62^{+1.6}_{-0.98}$   &    Margutti  \\ \hline
        24888    & 1213.1   & 17.85         & $2.53^{+1.57}_{-1.11}$ & 2.0 & $5.13^{+3.10}_{-2.25}$    &   Margutti \\ \hline
        24889    & 1214     & 16.86         & $2.01^{+1.47}_{-0.99}$ & 2.0 & $4.08^{+2.99}_{-2}$      &   Margutti \\ \hline
        23870    & 1250.1   & 28.52         & $1.17^{+0.74}_{-0.52}$ & 2.0 & $2.39^{+1.50}_{-1.08}$   &    Margutti  \\ \hline
        24923    & 1258     & 25.61         & $<$2.75                 & 2.0 & $<$5.56                 &    Margutti      \\ \hline
        24924    & 1258.7   & 19.82         & $<$1.99           & 2.0 & $<$4.03                       &    Margutti \\ \hline
        23869    & 1573.1   & 23.76         & $1.19^{+0.88}_{-0.60}$  & 2.1 & $2.51^{+1.86}_{-1.27}$   &   Margutti     \\ \hline
        26223    & 1574.6   & 29.69         & $<$3.4                  & 2.1 & $<$7.16                 &    Margutti \\ \hline
        24336    & 1576.2   & 38.56         & $0.38^{+0.67}_{-0.33}$ & 2.1 & $0.81^{+1.40}_{-0.70}$   &    Margutti     \\ \hline
        24337    & 1578     & 29.69         & $<$4.13                & 2.1 & $<$8.7                   &   Margutti        \\ \hline
        25733    & 1669.9   & 28.4          & $0.76^{+0.76}_{-0.46}$ & 2.1 & $1.64^{+1.60}_{-0.99}$   &    Troja   \\ \hline
        25734    & 1673.6   & 30.96         & $<$2.63               & 2.1 & $<$5.6                    &  Troja     \\ \hline
        25527    & 1733.8   & 36.58         & $<$2.23                & 2.1 & $<$4.77                  &   Troja        \\ \hline
        27088    & 1980.1   & 27.7          & $0.37^{+0.60}_{-0.30}$ & 2.2 & $0.82^{+1.38}_{-0.67}$   &  Troja      \\ \hline
        27089    & 1985.9   & 31.65         & $0.71^{+0.70}_{-0.40}$ & 2.2 & $1.59^{+1.50}_{-0.94}$     &  Troja \\ \hline
        27090    & 2000.6   & 29.49         & $0.37^{+0.60}_{-0.30}$ & 2.2 & $0.82^{+1.30}_{-0.65}$     &  Troja   \\ \hline
        27731    & 2022.1   & 14.87         & $<$5.44               & 2.2 & $<$12                     &  Troja           \\ \hline
        27091    & 2025     & 29.54         & $0.36^{+0.60}_{-0.28}$         & 2.2 & $0.8^{+1.31}_{-0.63}$               &    Troja  \\ \hline
        25528    & 2042.9   & 12.42         & $<$6.51               & 2.2 & $<$14.4                   &  Troja             \\ \hline
        27752    & 2043.26  & 12.4          & $<$3.83                  & 2.2   &  $<$8.47   &    Troja  \\ \hline
        27753    & 2043.57  & 12.41         & $0.92^{+1.40}_{-0.70}$ & 2.2 & $2.05^{+3.10}_{-1.50}$      &  Troja  \\ \hline
        27754    & 2043.88  & 12.41         & $<$6.51              & 2.2 & $<$14.4                    &  Troja               \\ \hline
        % 28525    & 2427.7   & 9.9           & $<$3.73                & 2.3 &  $<$8.58                       &  Troja           \\ \hline
        % 29370    & 2428     & 19.8          & $<$2.88                  & 2.3 &   $<$6.62                      &  Troja           \\ \hline
        % 28527    & 2432.2   & 14.88         &  $<$3.97             & 2.3 & $<$9.14                     &   Troja          \\ \hline
        % 29377    & 2434.6   & 15.3          & $0.75^{+1.16}_{-0.57}$  & 2.3 & $1.76^{+2.69}_{-1.32}$  &  Troja         \\ \hline
        % 28523    & 2440     & 15.3          & $0.72^{+1.13}_{-0.55}$  & 2.3    & $1.69^{+2.65}_{-1.30}$  &  Troja         \\ \hline
        % 29395    & 2440.4   & 15.17         & $<$3.89              & 2.3 &  $<$8.97                  &  Troja     \\ \hline
        % 28358    & 2449.2   & 10.4          & $1.11^{+1.70}_{-0.83}$                    & 2.3    &   $2.58^{+3.97}_{-1.94}$                         &  Troja    \\ \hline
        % 29397    & 2450.6   & 11.9          & $<$4.14                     & 2.3    &    $<$9.53                        &  Troja   \\ \hline
        % 29398    & 2451.2   & 21.78         & $0.51^{+0.80}_{-0.40}$                     & 2.3    & $1.18^{+1.86}_{-0.90}$                           &  Troja  \\ \hline
        % 29399    & 2451.7   & 23.1          & $0.49^{+0.77}_{-0.37}$                     & 2.3    &  $1.14^{+1.78}_{-0.87}$                          &  Troja  \\ \hline
        % 28526    & 2452.68  & 14.8          & $<$2.96                     & 2.3   &  $<6.82$                         &  Troja \\ \hline

\caption*{
    $^{a}$ Count rates are in units of $10^{-4}$ cts $s^{-1}$. All the values are corrected for PSF losses. \\
    $^{b}$ Fluxes in units of $10^{-15}$ erg $cm^{-2}$ $s^{-1}$. Values are corrected for Galactic extinction.  \\
    All data points have 1$\sigma$ errorbars; upper limits are 3$\sigma$.
}
   \end{longtable}

\end{Large}
%%%%%%%%%%%%%%%%%%%%%%%%%%%%%%%%%%%%%%%%%%%%%%%%%%

% Don't change these lines
\bsp	% typesetting comment
\label{lastpage}
\end{document}